\documentclass[final,3p,times,twocolumn, sort&compress]{elsarticle}

\usepackage{preprintcover}  

\PreprintCoverPaperTitle{Search for new phenomena
in photon+jet events collected in proton--proton
collisions at $\sqrt{s} = 8$~\tev{} with the ATLAS detector}
\PreprintIdNumber{CERN-PH-EP-2013-126}
\PreprintCoverAbstract{
This Letter describes a model-independent search for the production of
new resonances in
photon + jet (\gamjet{}) events using 20~\lumfb of proton--proton
LHC data recorded with the ATLAS detector at a centre-of-mass energy of $\sqrt{s} = 8$~\tev.  
The \gamjet{}~mass distribution is compared to a background model fit
from data; no significant deviation from the background-only
hypothesis is found. Limits are set at 95\% credibility level on generic
Gaussian-shaped signals and two benchmark phenomena beyond the
Standard Model: non-thermal quantum black holes and 
excited quarks.  Non-thermal quantum black holes are excluded below masses of
4.6~\tev\ and excited quarks are excluded below masses of 3.5~\tev.}
\PreprintJournalName{PLB}

\usepackage{subfig} 
\usepackage{subfloat}
\usepackage{mathrsfs}
\usepackage{graphicx}
\usepackage{xspace}
\usepackage{afterpage}
\usepackage{multirow}
\usepackage{dcolumn}
\usepackage{supertabular}
\usepackage[hyperref]{xcolor}
\usepackage{hyperref}
\usepackage{atlasphysics}

\newcommand{\gamjet}{\ensuremath{\gamma+\mathrm{jet}}} 
\newcommand{\Mgj}{\ensuremath{m_{\gamma j}}\xspace}
\newcommand{\qstar}{\ensuremath{q^{*}}\xspace}

\newcommand{\pb}{\ensuremath{\mathrm{pb}}\xspace}

\newcommand{\lumfb}{\ensuremath{\mathrm{fb}^{-1}}\xspace}

\newcommand{\pythia}{{\sc pythia}\xspace}

\newcommand{\sherpa}{{\sc sherpa}\xspace}
\newcommand{\jetphox}{{\sc jetphox}\xspace}

\begin{document}

\title{\boldmath
Search for new phenomena
in photon+jet events collected in proton--proton
collisions at $\sqrt{s}$ = 8~\tev{} with the ATLAS detector}

\begin{frontmatter}

\author{ATLAS Collaboration}
\begin{abstract}
This Letter describes a model-independent search for the production of
new resonances in
photon + jet (\gamjet{}) events using 20~\lumfb of proton--proton
LHC data recorded with the ATLAS detector at a centre-of-mass energy of $\sqrt{s} = 8$~\tev.  
The \gamjet{}~mass distribution is compared to a background model fit
from data; no significant deviation from the background-only
hypothesis is found. Limits are set at 95\% credibility level on generic
Gaussian-shaped signals and two benchmark phenomena beyond the
Standard Model: non-thermal quantum black holes and 
excited quarks.  Non-thermal quantum black holes are excluded below masses of
4.6~\tev\ and excited quarks are excluded below masses of 3.5~\tev.
\end{abstract}

\end{frontmatter}

\hyphenation{ATLAS}

\section{Introduction}
Several exotic production mechanisms have been proposed that produce
massive photon + jet (\gamjet) final states. They include non-thermal quantum black holes
(QBHs)~{\cite{Meade:2007sz,Calmet:2008dg,Gingrich:2009hj}}, 
excited quarks~\cite{Bhattacharya:2009xg,PhysRevD42815,PhysRevD.60.034004},
quirks~\cite{Kang:2008ea,Martin:2010kk,Okun19801}, Regge excitations
of string
theory~\cite{Anchordoqui:2008ac,Anchordoqui:2008di,Nath2010185},
and topological pions~\cite{Bai:2010mn}.  Of the past
searches~\cite{PhysRevLett.72.3004,adloff-2000-17,zeus-collaboration-2002-549,Affolder:2001yp,ATLAS:2011ai},
the only LHC search for this signature was done using proton--proton ($pp$) collision
data obtained at a centre-of-mass energy of $\sqrt{s} = 7$~\tev\ with
the ATLAS detector. It found no evidence of new physics and placed
upper limits on the visible signal cross-section in the range 1.5--100~fb
and excluded excited-quark masses up to 2.46~\tev~at the 95\% credibility
level (CL)~\cite{ATLAS:2011ai}. The present Letter describes a
model-independent search for $s$-channel \gamjet\ production, improved
over the earlier search. It presents the first limits on QBHs decaying
to the \gamjet{} final state and places new limits both on
excited quarks
and on generic Gaussian-shaped sources which describe other
narrow resonant signals such as topological pions. Sensitivity to such
signals has been improved compared to the previous search through a
combination of an order-of-magnitude larger data sample
(20.3~fb$^{-1}$), a higher centre-of-mass energy ($\sqrt{s} =
8$~\tev{}), reduced background uncertainties, and improved selection
criteria at high invariant mass.

The Standard Model (SM) of particle physics lacks a mechanism whereby $pp$
collisions produce resonances that subsequently decay to a \gamjet\ final state.
Direct \gamjet~ production can occur at tree
level via Compton scattering of a quark and a gluon, or through
quark--antiquark annihilation.  The former process accounts for most of the
direct \gamjet~ production.
Events
with a high transverse momentum photon and one or more jets can also
arise from radiation off final-state quarks, or from dijet or multi-jet
processes, where 
secondary photons, referred to as fragmentation photons, are produced
during fragmentation
of the hard-scattered quarks or
gluons~\cite{Catani:2002ny,Aurenche:2006vj,Belghobsi:2009hx,Ichou:2010wc}.
The \gamjet~ invariant mass (\Mgj) distribution
resulting from this mixture of processes is smooth and rapidly
falling,  and is therefore well suited to revealing high-mass
resonances decaying to \gamjet.

The \Mgj\ distribution is used to search for a peak over the
SM background, estimated by fitting a smoothly falling function to the
\Mgj~distribution in the region \Mgj $>$ 426 \gev.  In the absence of
a signal, Bayes' theorem is used to set limits on Gaussian-shaped
signals and on two benchmark models: QBHs and excited quarks.

Models with extra dimensions, such as the
Arkani-Hamed--Dimopoulous--Dvali (ADD)
model~\cite{ArkaniHamed:1998rs,Antoniadis:1998ig},
solve the mass hierarchy problem of the SM by lowering the fundamental
scale of quantum gravity ($M_{D}$) to a few~\tev. Consequently, the
LHC could produce quantum black holes with masses at or above
$M_{D}$~\cite{Dimopoulos:2001hw,Giddings:2001bu}. QBHs produced near
$M_{D}$ would evaporate faster than they thermalize, decaying into a few
particles rather than high-multiplicity final
states~\cite{Calmet:2008dg,Gingrich:2009hj}.
Regardless of the number of extra dimensions $n$, such a signal would
appear as a
local excess over the steeply falling $m_{\gamma j}$ distribution near
the threshold mass ($M_\mathrm{th}$)
and would fall 
exponentially at higher masses.
Searches performed by the CMS Collaboration for QBHs with
high-multiplicity
energetic final states yielded limits in the range of $4.3$--$6.2$
\tev, for $n=1$--$6$ and different model
assumptions~\cite{Chatrchyan:2013xva}.
 This Letter
assumes $M_\mathrm{th} = M_{D}$
and $n = 6$, where the cross-section times branching fraction for QBH production and
decay to \gamjet\ final states at $M_\mathrm{th} =$ 1, 3 and 5 \tev\
is 200, 0.3 and $6 \times 10^{-5}$ \pb, respectively~\cite{Gingrich:2009hj}.
For decays to dijet final states at these same threshold masses, the rates are larger by factors of 11, 39 and 125.

Excited-quark ($q^*$) states, which the ATLAS and CMS experiments have
also sought in dijet final
states~\cite{ATLAS:2012pu,Chatrchyan:2013qha,CMS:2012yf}, could be produced via
the fusion of a gluon with a quark.
The model is defined by one
parameter, the excited-quark mass $m_{q^*}$, with the compositeness
scale set to $m_{q^*}$. 
Only gauge interactions are considered with
the SU(3), SU(2), and U(1) coupling
multipliers fixed to $f_\mathrm{s} = f = f^\prime = 1$~\cite{PhysRevD42815}.  
This results in branching fractions
for $q^{*} \rightarrow qg$ and $q^{*} \rightarrow q \gamma$ of 0.85
(0.85) and 0.02 (0.005), respectively, for $q=u$ ($q=d$). 
The leading-order cross-sections times branching fractions
combining all flavours of excited quarks for $m_{q^{*}} = $ 1, 3 and 5
\tev\ are 4, $2 \times 10^{-3}$ and $3 \times 10^{-6}$ \pb, respectively.

Factorization and renormalization scale uncertainties are not used
for either signal type, for comparison with earlier
analyses~\cite{ATLAS:2011ai,ATLAS:2012pu,Chatrchyan:2013qha}.

\section{Signal and background simulation samples}
To cross-check the data-driven background estimates, the SM prompt
photon processes are simulated with \pythia~8.165~\cite{Pythia:2007} and
\sherpa~1.4.0~\cite{Gleisberg:2008ta}.  The \pythia\ and \sherpa\
prompt photon samples use CTEQ6L1~\cite{Pumplin:2002vw} and
CT10~\cite{Lai:2010vv} leading-order and next-to-leading-order parton distribution functions (PDFs),
respectively.  The simulated samples of QBHs are obtained from the {\sc
qbh~1.05} generator~\cite{Gingrich:2009da} followed by parton
showering using \pythia~8.165.  The simulated $q^*$ signal samples are
generated with the excited-quark model in \pythia~8.165.  Both signal
generators use the MSTW2008LO~\cite{Martin:2009iq} leading-order PDF
set with the AU2 underlying-event tune~\cite{atlas:pythia8tunes}.
Additional inelastic $pp$ interactions, termed pileup, are included in
the event simulation by overlaying simulated minimum bias events with
an average of 20 interactions per bunch crossing. 
All the above Monte Carlo (MC) simulated samples
are produced using the ATLAS full {\sc
geant4}~\cite{Agostinelli2003250} detector
simulation~\cite{ATLASSimInfra}.  Supplementary studies of the
background shape are also performed with the next-to-leading-order
\jetphox~1.3.0
generator~\cite{Catani:2002ny,Belghobsi:2009hx,Aurenche:2006vj} at
parton level using CT10 PDFs.

\section{The ATLAS detector}
A detailed description of the detector is available in
Ref.~\cite{1748-0221-3-08-S08003}, and the event selection is similar
to that described in Ref.~\cite{ATLAS:2011ai}.
Photons are detected by a lead--liquid-argon sampling
electromagnetic calorimeter (EMC).
The EMC has a pre-sampler layer and three additional, differently
segmented, layers; only the
first two are used in photon identification.
Upstream of the EMC, the inner detector allows an accurate
reconstruction of tracks from the primary $pp$ collision point and also
from secondary vertices, permitting an efficient reconstruction of
photon conversions in the inner detector.
For $|\eta| < 1.37$~\footnote{ATLAS uses a right-handed coordinate
system with its origin at the nominal interaction point (IP) in the
centre of the detector and the $z$-axis along the beam pipe. The
$x$-axis points from the IP to the centre of the LHC ring, and the
$y$-axis points upward. Cylindrical coordinates $(r,\phi)$ are used
in the transverse plane, $\phi$ being the azimuthal angle around the
beam pipe. The pseudorapidity is defined in terms of the polar angle
$\theta$ as $\eta=-\ln\tan(\theta/2)$.} 
an iron--scintillator tile calorimeter behind the EMC provides hadronic
coverage.
The endcap and forward regions, $1.5 < |\eta| < 4.9$, are instrumented
with liquid-argon calorimeters for both the electromagnetic and hadronic 
measurements. 
Events for this analysis were collected with a trigger requiring at least one photon
candidate with transverse momentum (\pt) above 120~\gev~\cite{Aad:2012xs}.
The integrated luminosity of the data sample\footnote{The systematic uncertainty
  on the luminosity is derived, following the same methodology as that
  detailed in Ref.~\cite{ref:lumi},  from a preliminary calibration of the luminosity scale derived from beam-separation scans performed in November 2012.}
 is $(20.3\pm
0.6)$~\lumfb.

\section{Event selection}
Each event is required to contain a primary vertex with at least two
tracks each with $\pt > 400$~\mev.  If more than one vertex is found, the
primary vertex is defined as the one with the highest scalar summed 
\ptsq\ of associated tracks.

Jets are reconstructed from clusters of calorimeter cells~\cite{springerlink:10.1140/epjc/s10052-010-1512-2}, using the
anti-$k_t$ clustering algorithm~\cite{Cacciari:2008gp} with radius
parameter $R = 0.6$. The effects on jet energies due to multiple $pp$
collisions in the same or in neighbouring bunch crossings are accounted
for by a jet-area-based
correction~\cite{Cacciari:2008gn,Cacciari:2009dp}.  Jet energies are
calibrated to the hadronic energy scale using corrections from MC simulation and the combination
of several in situ techniques applied to data~\cite{2010JES}.  Events
are discarded if the leading (highest-\pt)  jet is affected by noise or hardware problems in the detector, or is identified as arising from non-collision backgrounds.
Only jets with $|\eta_j|<2.8$ are
considered further.

Photon candidates are reconstructed from clusters in the
electromagnetic calorimeter and tracking information provided by the
inner detector. Inner detector tracking information is used to reject
electrons and to recover photons converted to $e^+e^-$
pairs~\cite{Aad:2011tw}. 
Photon candidates satisfy standard ATLAS selection
criteria that are designed to reject backgrounds from
hadrons~\cite{PhysRevD.83.052005}. The photon candidates must meet
$\eta$-dependent requirements on hadronic leakage and shower shapes in
the first two sampling layers of the electromagnetic calorimeter.
Energy calibrations are applied to photon candidates to account for
energy loss upstream of the electromagnetic calorimeter and for both
lateral and longitudinal shower leakage. The simulation is corrected for
differences between data and MC events for each photon shower shape
variable. Events are discarded if the leading photon is reconstructed using
calorimeter cells affected by noise bursts or transient hardware
problems.

These photon identification criteria reduce instrumental backgrounds
to a negligible level, but some background from fragmentation photons
and hadronic jets remains. This background is further reduced by
requirements on nearby calorimeter activity.  Energy deposited in the
calorimeter near the photon candidate, $E_\mathrm{T}^\mathrm{isol}$,
must be no larger than $0.011\ p_\mathrm{T}^\gamma + 3.65$~GeV, a
criterion that provides constant efficiency for all pileup conditions
and over the entire \pt\ range explored. This transverse isolation energy is
calculated by summing the energy as measured in electromagnetic and
hadronic calorimeter cells inside a cone of radius $\Delta R =
\sqrt{(\Delta\eta)^2 + (\Delta\phi)^2} = 0.4$ centred on the photon
cluster, but excluding the energy of the photon cluster itself, and is
corrected on an event-by-event basis for the ambient energy density
due to pileup and the underlying event, as well as energy leakage from
the photon cluster into the cone. Additionally, the photon is required
to have angular separation of $\Delta R(\gamma,\mathrm{jet}) > 1.0$ 
between the leading photon and all other
jets with $p_\mathrm{T} > 30$~\gev, with the exception of a required
photon-matched jet.
Such photon-matched jets arise  from the fact that photon energy deposits in the calorimeter are also
reconstructed as jets. To further suppress background 
from fragmentation photons, where the angular separation between the
photon and the corresponding photon-matched jet can
be large, the leading photon candidate is required to 
have exactly one reconstructed jet with $\Delta R(\gamma, \mathrm{jet}) < 0.1$.
This photon-matched jet is not considered in any other selection criteria, including those related to photon isolation.

Events containing at least one photon candidate and
at least one jet candidate, each with $\pt>125$~\gev{}, are selected for
final analysis. The photon trigger is fully efficient for these events. 
In the events where more than one photon or jet is found, the
highest-\pt\ candidates are selected to constitute the photon and
jet pair to compute \Mgj.

The sensitivity of the search is improved by
requirements on photon and jet pseudorapidities.
Dijet production rates increase with jet  absolute pseudorapidity whereas rates
for an $s$-channel signal would diminish.
Photons are required to be in the barrel calorimeter, $|\eta_\gamma| <
1.37$, and the distance between the photon and jet, $\Delta\eta =
|\eta_\gamma - \eta_j|$, must be less than 1.6.
The latter requirement was chosen by optimizing the expected significance of signals, using
the $\Delta\eta$ distribution found in QBH
and excited-quark signal simulations, with respect to the SM
background as predicted by the \pythia\ prompt photon
simulation. 

The acceptance of the event selection is about 60\%.  It is calculated
using parton-level quantities by imposing the kinematic selection criteria
 (photon/jet $|\eta|$, photon/jet
\pt,  $\Delta\eta$, $\Delta R$).   All other
selections, which in general correspond to event and object quality
criteria, were used to calculate the efficiency based on the events included in the acceptance.  The 
efficiency falls from 83\% to 72\% for masses from 1 \tev{} to 6
\tev\ for QBH signals and from 85\% to 80\% for excited-quark signals over the same mass range.
There are 285356 events in the data sample after all event selections. The 
highest \Mgj\ value observed is 2.57 \tev.

\section{Background estimation}
The combined SM and instrumental background to the search is
determined by fitting the \Mgj\ distribution to the four-parameter ansatz
function~\cite{Harris:2011bh},
\begin{equation} \label{eq1}
f(x \equiv m_{\gamma j}/\sqrt{s}) = p_1 (1-x)^{p_2}x^{-(p_3 +
p_4\ln x)}\, . 
\end{equation}
\noindent
The functional form has been tested with \pythia\ and \sherpa\
prompt photon simulations and next-to-leading-order \jetphox
predictions with comparable sample size. Two additional control
samples in the data are also defined to further validate the
functional form. The first control sample is defined by
reversing two of the photon identification criteria, $\Delta E$
and $E_{\mathrm{ratio}}$~\cite{PhysRevD.83.052005}, that compare
the lateral shower shapes of single photons in the first layer of the
calorimeter to those of jets with high electromagnetic energy fraction
and low particle multiplicity, typical for meson decays.
This sample has a similar \Mgj\ shape to the dominant background, 
SM \gamjet\ events. 
The second control sample is defined by reversing the
photon isolation criterion, $E_\mathrm{T}^\mathrm{isol}$. This control
sample is enriched in the second largest background, dijet events
in which a jet has passed the photon identification cuts.

\begin{figure}[htb]
\centering
\includegraphics[width=\columnwidth]{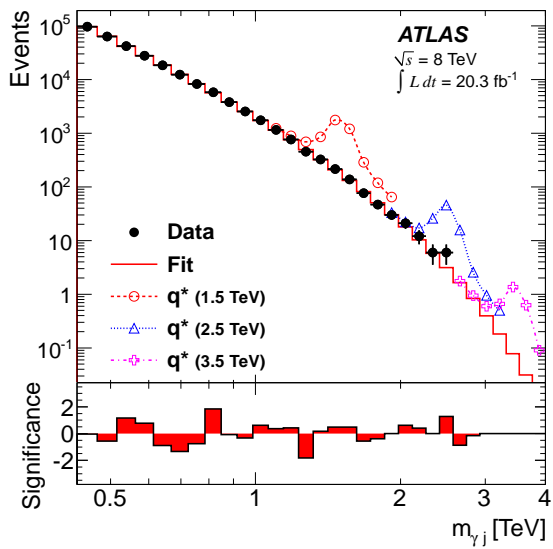}
\caption{Invariant mass of the \gamjet\  pair for
events passing the final selections.  The bin widths are chosen to be twice the mass resolution at the centre of each bin.
Overlaid is the fitted background function integrated over each bin
(solid line), with three examples of \qstar signals, as described in the text.
For
better visibility 
the \qstar signals are only drawn for \Mgj{} within $\pm$25\% of the nominal signal mass. 
The bottom panel shows the statistical significance of the difference
between data and background in each bin. }
\label{fig1}
\end{figure}

Fig.~\ref{fig1} shows the resulting distribution of the \gamjet\
invariant mass. The bin widths are chosen to be twice the mass
resolution at the centre of each bin.
The relative resolution
is about 4\% of $m_{\gamma j}$ at 1~\tev{}, improving to about 3\% at
2~\tev{}.
The fit result is also shown in
Fig.~\ref{fig1}.  The bottom panel of the figure shows the statistical
significance of the difference between data and the fit in each
bin~\cite{Choudalakis11b}.  The fit quality is quantified using a
negative log-likelihood test statistic.  The probability of the fit
quality to be at least as good as the observed fit ($p$-value) is
74\%, indicating that the data are consistent with the functional
form. 

\section{Results}
\subsection{Search results}
The search region is defined to be \Mgj\ $>$ 426 \gev, which is the
lower edge of the first bin for
which biases due to kinematic and trigger threshold effects are negligible.
The \gamjet\ search is sensitive to new resonances in the region
between 426~\gev{} and 1~\tev{}, 
where the statistics of dijet searches are limited 
by the higher hadronic trigger thresholds.
The {\sc bumphunter} algorithm~\cite{Choudalakis11a} is used to search
for  statistical evidence of a~resonance.
The algorithm operates on the binned $m_{\gamma j}$ distribution,
comparing the background estimate with the data in mass intervals of
varying numbers of adjacent bins across the entire distribution. 
For each interval in the scan, it computes the significance of any excess
found.
The significance of the outcome is evaluated using the ensemble of
possible outcomes in any part of  the distribution 
under the background-only hypothesis, obtained by repeating the analysis on
pseudodata drawn from the background function.  The algorithm identifies the two-bin interval 785--916~\gev\ as the single most
discrepant  interval.
Before including systematic uncertainties, the $p$-value
is 61\%, including the trials factor, or ``look-elsewhere'' effect.
Thus, the excess is not significant and the data are consistent 
with a smoothly falling background.

\subsection{Limit results}
In the absence of any signal, three types of \gamjet{} signals are
explored: a generic Gaussian-shaped signal with an arbitrary production
cross-section, resulting from resonances with 
varying intrinsic widths convolved with the detector resolution; the
QBH model; and the
excited-quark model.  For each signal mass considered, 
the fit to the observed mass distribution is repeated with the sum of
the four-parameter background function (Eq. (\ref{eq1})) and a signal
template with a normalization determined during the fit.
Bayesian limits at the 95\% CL are computed as described
in Ref.~\cite{ATLAS:2012pu} using a prior probability density that is
constant for positive values of the signal production cross-section
and zero for unphysical, negative values.

Systematic uncertainties affecting the limits on production of new
signals are evaluated.  The signal yield is subject to systematic
uncertainties on the integrated luminosity (2.8\%), photon isolation
efficiency (1.2\%), trigger efficiency (0.5\%), and photon
identification efficiencies (1.5\%). The last of these includes
extrapolation to high \pt~(0.1\%) and pileup effects (0.1\%).
Uncertainties on the jet and photon energy scale contribute 1.0~--~1.5\% and 0.3\%, respectively, through their effects on the shape and yield of
the signal distribution. The sizes of the systematic uncertainties are
similar for the $q^*$ and QBH signals.  These systematic
uncertainties are treated as marginalized nuisance parameters in the
limit calculation. 
Systematic uncertainties on the value and shape of the	
signal acceptance due to the PDF uncertainties were examined and found to be negligible.
To account for the statistical uncertainties
on the background fit parameters, the background function is
repeatedly fit to pseudodata for which the content of each bin is drawn from
Poisson distributions.  The mean of the Poisson distribution for a
given bin corresponds to the number of entries actually observed in
that bin in the data.  The variations in the fit predictions for a~given bin, 1\% of the background at 1~\tev\ to about 20\% of the
background at 3~\tev, are taken as indicative of the systematic
uncertainty.  This bin-by-bin uncertainty is treated in the limit as
fully correlated, using a single nuisance parameter that scales the
entire background distribution. Several other fit functions from
Ref.~\cite{Harris:2011bh} were tested, and a negligible systematic
uncertainty was found. 

\begin{figure}[htb]
\centering
\includegraphics[width=\columnwidth]{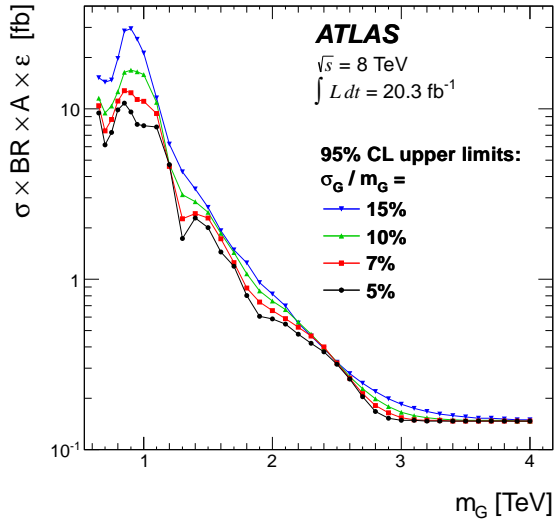}
\caption{The 95\% CL upper limits on $\sigma\times
BR\times A\times \varepsilon$ for a hypothetical signal with a Gaussian-shaped
$m_{\gamma j}$ distribution as a function
of the signal mass $m_\mathrm{G}$ for four values of the relative width $\sigma_\mathrm{G}/m_\mathrm{G}$.} 
\label{fig2}
\end{figure}

Fig.~\ref{fig2} shows the model-independent limits on the
visible
cross-section, defined as the product of the cross-section ($\sigma$)
times branching fraction (BR) times acceptance ($A$) times efficiency
($\varepsilon$), of a potential signal as a function of the mass of
each signal template, and includes the systematic uncertainties
discussed above.  The signal line shape is modelled as a Gaussian
distribution, with one of four relative widths:
$\sigma_\mathrm{G}/m_\mathrm{G} = 5\%,
7\%$, 10\%, and 15\%, where $\sigma_\mathrm{G}$ ($m_{\mathrm{G}}$) is
the width (mean mass)
of the Gaussian.  
The differences between the limits for
different widths are driven by the increased sensitivity to local
fluctuations for the narrower signals. 
Beyond the highest-mass event recorded, 2.57~\tev, the limits begin to converge due to the
absence of observed events.  At 1 \tev\ and 4 \tev\ the limits are 8~fb
and 0.1~fb, respectively,
for $\sigma_\mathrm{G}/m_\mathrm{G}$ = 5\%. 
At 3~\tev, the new limit
improves  the earlier ATLAS result in this channel by an order of
magnitude. 

\begin{figure}[htb]
\centering
\includegraphics[width=\columnwidth]{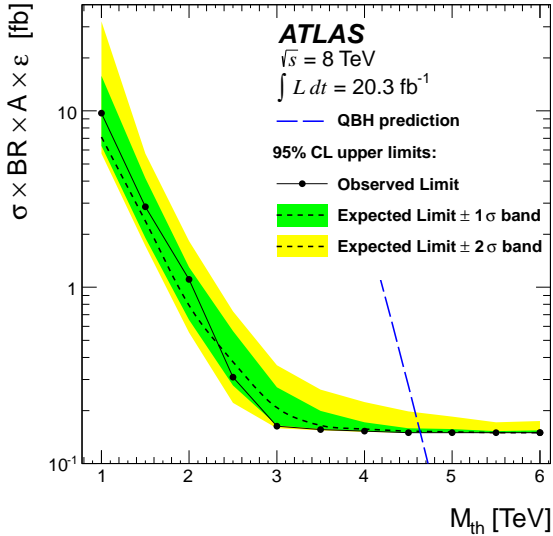}
\caption{The 95\% CL upper limits on $\sigma \times
BR\times A\times \varepsilon$ for QBHs decaying to a
photon and a jet, as a function of the threshold mass $M_\mathrm{th}$,
assuming  $M_{D} = M_\mathrm{th}$ and $n=6$.
The limits take into account statistical and systematic uncertainties. 
Points along the solid black line indicate the mass of the signal
where the limit is computed.
The black short dashed line is the central value of the expected limit. 
Also shown are the $\pm 1\sigma$ and $\pm 2\sigma$ uncertainty bands
indicating the underlying distribution of possible limit outcomes under the
background-only hypothesis.
The predicted visible cross-section for QBHs is shown as the long dashed line.}
\label{fig3A}
\end{figure}
The limit on the visible cross-section in the QBH
model is shown in Fig.~\ref{fig3A} as a function of 
$M_{\mathrm{th}}$.
The observed (expected) lower limit
on the QBH mass threshold is found to be 4.6 (4.6) \tev, at 95\% CL.
The uncertainty on the QBH theoretical cross-section
  arising from PDF uncertainties moves the uppermost excluded mass by 0.2\%.

\begin{figure}[htb]
\centering
\includegraphics[width=\columnwidth]{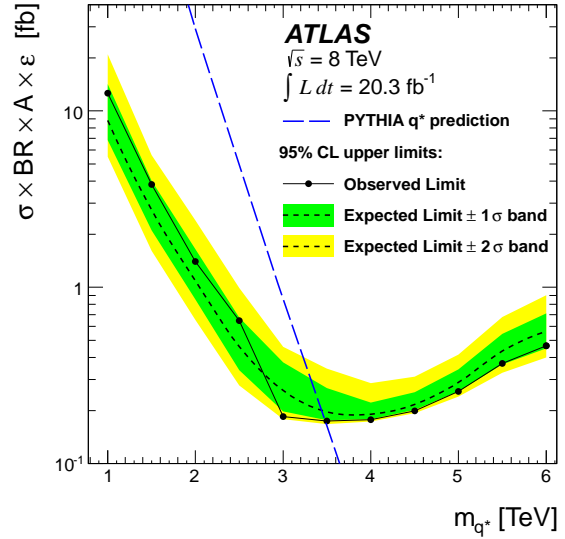}
\caption{The 95\% CL upper limits on $\sigma \times
BR\times A\times \varepsilon$ for 
excited quarks decaying to a
photon and a jet, as a function of the signal mass $m_{q^*}$.
The limits take into account statistical and systematic uncertainties. 
Points along the solid black line indicate the mass of the signal
where the limit is computed.
The black short dashed line is the central value of the expected limit. 
Also shown are the $\pm 1\sigma$ and $\pm 2\sigma$ uncertainty bands
indicating the underlying distribution of possible limit outcomes under the
background-only hypothesis.
The long dashed line shows the predicted visible cross-section for excited-quark production from \pythia.}
\label{fig3B}
\end{figure}
The limit on the visible cross-section
in the excited-quark model as a function of the $q^*$ mass, assumed
to be the same for $u^{*}$ and $d^{*}$,  is shown in
Fig.~\ref{fig3B}.
The rise in the expected and observed limits at high
$m_{q^*}$ is due
to the increased fraction of off-shell production of the \qstar,
which alters the signal distribution to lower masses with a wider peak. 
The observed
(expected) lower limit on the excited-quark mass is found to be 3.5
(3.4) \tev, at 95\% CL. With a much lower branching fraction than
the dijet channel but also smaller backgrounds, this result improves
on the present exclusion limits in the dijet final state: 3.32~\tev\
from CMS with 5~\lumfb\ of data at $\sqrt{s} = 7$ \tev{}~\cite{CMS:2012yf},
and 2.83~\tev\ from ATLAS with
4.8~\lumfb~\cite{ATLAS:2012pu} of data at $\sqrt{s} = 7$ \tev{}.
The uncertainty on the \qstar theoretical cross-section
  arising from PDF uncertainties moves the uppermost excluded mass by 0.9\%.

\section{Conclusions}
In conclusion, the \gamjet~mass distribution measured in
20.3~\lumfb\ of $pp$ collision data, collected at $\sqrt{s} = 8$ \tev{} by the ATLAS experiment at the LHC,
is well described by the background model and no evidence for new
phenomena is found.
Limits at 95\%~CL using Bayesian statistics are presented for signal processes
yielding a Gaussian line shape, non-thermal
quantum black holes, and excited quarks.
The limits on Gaussian-shaped resonances exclude 4~\tev\ resonances
with visible cross-sections near 0.1~fb.
Non-thermal quantum black hole and excited-quark models with a 
\gamjet~final state are excluded for masses up to 4.6~\tev\ and 3.5~\tev,
respectively.
The limits reported here on the production of new resonances in
the \gamjet~final state are the most stringent limits set to date
in this channel.

We thank CERN for the very successful operation of the LHC, as well as the
support staff from our institutions without whom ATLAS could not be
operated efficiently.

\section*{Acknowledgments}
We acknowledge the support of ANPCyT, Argentina; YerPhI, Armenia; ARC,
Australia; BMWF and FWF, Austria; ANAS, Azerbaijan; SSTC, Belarus; CNPq and FAPESP,
Brazil; NSERC, NRC and CFI, Canada; CERN; CONICYT, Chile; CAS, MOST and NSFC,
China; COLCIENCIAS, Colombia; MSMT CR, MPO CR and VSC CR, Czech Republic;
DNRF, DNSRC and Lundbeck Foundation, Denmark; EPLANET, ERC and NSRF, European Union;
IN2P3-CNRS, CEA-DSM/IRFU, France; GNSF, Georgia; BMBF, DFG, HGF, MPG and AvH
Foundation, Germany; GSRT and NSRF, Greece; ISF, MINERVA, GIF, DIP and Benoziyo Center,
Israel; INFN, Italy; MEXT and JSPS, Japan; CNRST, Morocco; FOM and NWO,
Netherlands; BRF and RCN, Norway; MNiSW, Poland; GRICES and FCT, Portugal; MERYS
(MECTS), Romania; MES of Russia and ROSATOM, Russian Federation; JINR; MSTD,
Serbia; MSSR, Slovakia; ARRS and MVZT, Slovenia; DST/NRF, South Africa;
MICINN, Spain; SRC and Wallenberg Foundation, Sweden; SER, SNSF and Cantons of
Bern and Geneva, Switzerland; NSC, Taiwan; TAEK, Turkey; STFC, the Royal
Society and Leverhulme Trust, United Kingdom; DOE and NSF, United States of
America.

The crucial computing support from all WLCG partners is acknowledged
gratefully, in particular from CERN and the ATLAS Tier-1 facilities at
TRIUMF (Canada), NDGF (Denmark, Norway, Sweden), CC-IN2P3 (France),
KIT/GridKA (Germany), INFN-CNAF (Italy), NL-T1 (Netherlands), PIC (Spain),
ASGC (Taiwan), RAL (UK) and BNL (USA) and in the Tier-2 facilities
worldwide.

\bibliographystyle{atlasBibStyleWoTitle}
\bibliography{paper}

\providecommand{\href}[2]{#2}\begingroup\raggedright\begin{thebibliography}{10}

\bibitem{Meade:2007sz}
P.~Meade and L.~Randall,
  \href{http://dx.doi.org/10.1088/1126-6708/2008/05/003}{JHEP {\bfseries 0805}
  (2008) 003},
\href{http://arxiv.org/abs/0708.3017}{{\ttfamily arXiv:0708.3017 [hep-ph]}}.

\bibitem{Calmet:2008dg}
X.~Calmet, W.~Gong, and S.~D. Hsu,
  \href{http://dx.doi.org/10.1016/j.physletb.2008.08.011}{Phys. Lett. B
  {\bfseries 668} (2008)},
\href{http://arxiv.org/abs/0806.4605}{{\ttfamily arXiv:0806.4605 [hep-ph]}}.

\bibitem{Gingrich:2009hj}
D.~M. Gingrich, \href{http://dx.doi.org/10.1088/0954-3899/37/10/105008}{J.
  Phys. G {\bfseries 37} (2010) 105008},
  \href{http://arxiv.org/abs/0912.0826}{{\ttfamily arXiv:0912.0826 [hep-ph]}}.

\bibitem{Bhattacharya:2009xg}
S.~Bhattacharya, S.~S. Chauhan, B.~C. Choudhary, and D.~Choudhury,
  \href{http://dx.doi.org/10.1103/PhysRevD.80.015014}{Phys. Rev. D {\bfseries
  80} (2009) 015014},
\href{http://arxiv.org/abs/0901.3927}{{\ttfamily arXiv:0901.3927 [hep-ph]}}.

\bibitem{PhysRevD42815}
U.~Baur, M.~Spira, and P.~M. Zerwas,
  \href{http://dx.doi.org/10.1103/PhysRevD.42.815}{Phys. Rev. D {\bfseries 42}
  (1990) 815}.

\bibitem{PhysRevD.60.034004}
O.~\c{C}akir and R.~Mehdiyev,
  \href{http://dx.doi.org/10.1103/PhysRevD.60.034004}{Phys. Rev. D {\bfseries
  60} (1999) 034004}.

\bibitem{Kang:2008ea}
J.~Kang and M.~A. Luty,
  \href{http://dx.doi.org/10.1088/1126-6708/2009/11/065}{JHEP {\bfseries 0911}
  (2009) 065}, \href{http://arxiv.org/abs/0805.4642}{{\ttfamily arXiv:0805.4642
  [hep-ph]}}.

\bibitem{Martin:2010kk}
S.~P. Martin, \href{http://dx.doi.org/10.1103/PhysRevD.83.035019}{Phys. Rev. D
  {\bfseries 83} (2011) 035019},
  \href{http://arxiv.org/abs/1012.2072}{{\ttfamily arXiv:1012.2072 [hep-ph]}}.

\bibitem{Okun19801}
L.~B. Okun, \href{http://dx.doi.org/10.1016/0550-3213(80)90439-3}{Nucl. Phys. B
  {\bfseries 173} (1980) 1}.

\bibitem{Anchordoqui:2008ac}
L.~A. Anchordoqui, H.~Goldberg, S.~Nawata, and T.~R. Taylor,
  \href{http://dx.doi.org/10.1103/PhysRevD.78.016005}{Phys. Rev. D {\bfseries
  78} (2008) 016005}, \href{http://arxiv.org/abs/0804.2013}{{\ttfamily
  arXiv:0804.2013 [hep-ph]}}.

\bibitem{Anchordoqui:2008di}
L.~A. Anchordoqui {et~al.},
  \href{http://dx.doi.org/10.1103/PhysRevLett.101.241803}{Phys. Rev. Lett.
  {\bfseries 101} (2008) 241803},
  \href{http://arxiv.org/abs/0808.0497}{{\ttfamily arXiv:0808.0497 [hep-ph]}}.

\bibitem{Nath2010185}
P.~Nath {et~al.},
  \href{http://dx.doi.org/10.1016/j.nuclphysbps.2010.03.001}{Nucl. Phys. B,
  Proc. Suppl. {\bfseries 200-202} (2010) 185}.

\bibitem{Bai:2010mn}
Y.~Bai and A.~Martin,
  \href{http://dx.doi.org/10.1016/j.physletb.2010.08.058}{Phys. Lett. B
  {\bfseries 693} (2010) 292}, \href{http://arxiv.org/abs/1003.3006}{{\ttfamily
  arXiv:1003.3006 [hep-ph]}}.

\bibitem{PhysRevLett.72.3004}
{CDF} Collaboration, F.~Abe {et~al.},
  \href{http://dx.doi.org/10.1103/PhysRevLett.72.3004}{Phys. Rev. Lett.
  {\bfseries 72} (1994) 3004}.

\bibitem{adloff-2000-17}
{H1} Collaboration, C.~Adloff {et~al.},
  \href{http://dx.doi.org/10.1007/s100520000503}{Eur. Phys. J. C {\bfseries 17}
  (2000) 567}, \href{http://arxiv.org/abs/hep-ex/0007035}{{\ttfamily
  arXiv:hep-ex/0007035 [hep-ex]}}.

\bibitem{zeus-collaboration-2002-549}
{ZEUS} Collaboration, S.~Chekanov {et~al.},
  \href{http://dx.doi.org/10.1016/S0370-2693(02)02863-0}{Phys. Lett. B
  {\bfseries 549} (2002) 32},
  \href{http://arxiv.org/abs/hep-ex/0109018}{{\ttfamily arXiv:hep-ex/0109018
  [hep-ex]}}.

\bibitem{Affolder:2001yp}
{CDF} Collaboration, T.~Affolder {et~al.},
  \href{http://dx.doi.org/10.1103/PhysRevD.65.052006}{Phys. Rev. D {\bfseries
  65} (2002) 052006}, \href{http://arxiv.org/abs/hep-ex/0106012}{{\ttfamily
  arXiv:hep-ex/0106012 [hep-ex]}}.

\bibitem{ATLAS:2011ai}
{ATLAS Collaboration},
  \href{http://dx.doi.org/10.1103/PhysRevLett.108.211802}{Phys. Rev. Lett.
  {\bfseries 108} (2012) 211802},
  \href{http://arxiv.org/abs/1112.3580}{{\ttfamily arXiv:1112.3580 [hep-ex]}}.

\bibitem{Catani:2002ny}
S.~Catani, M.~Fontannaz, J.~Guillet, and E.~Pilon,
  \href{http://dx.doi.org/10.1088/1126-6708/2002/05/028}{JHEP {\bfseries 0205}
  (2002) 028},
\href{http://arxiv.org/abs/hep-ph/0204023}{{\ttfamily arXiv:hep-ph/0204023
  [hep-ph]}}.

\bibitem{Aurenche:2006vj}
P.~Aurenche {et~al.}, \href{http://dx.doi.org/10.1103/PhysRevD.73.094007}{Phys.
  Rev. D {\bfseries 73} (2006) 094007},
\href{http://arxiv.org/abs/hep-ph/0602133}{{\ttfamily arXiv:hep-ph/0602133
  [hep-ph]}}.

\bibitem{Belghobsi:2009hx}
Z.~Belghobsi {et~al.},
  \href{http://dx.doi.org/10.1103/PhysRevD.79.114024}{Phys. Rev. D {\bfseries
  79} (2009) 114024},
\href{http://arxiv.org/abs/0903.4834}{{\ttfamily arXiv:0903.4834 [hep-ph]}}.

\bibitem{Ichou:2010wc}
R.~Ichou and D.~d'Enterria,
  \href{http://dx.doi.org/10.1103/PhysRevD.82.014015}{Phys. Rev. D {\bfseries
  82} (2010) 014015},
\href{http://arxiv.org/abs/1005.4529}{{\ttfamily arXiv:1005.4529 [hep-ph]}}.

\bibitem{ArkaniHamed:1998rs}
N.~Arkani-Hamed, S.~Dimopoulos, and G.~Dvali,
  \href{http://dx.doi.org/10.1016/S0370-2693(98)00466-3}{Phys. Lett. B
  {\bfseries 429} (1998) 263},
\href{http://arxiv.org/abs/hep-ph/9803315}{{\ttfamily arXiv:hep-ph/9803315
  [hep-ph]}}.

\bibitem{Antoniadis:1998ig}
I.~Antoniadis, N.~Arkani-Hamed, S.~Dimopoulos, and G.~Dvali,
  \href{http://dx.doi.org/10.1016/S0370-2693(98)00860-0}{Phys. Lett. B
  {\bfseries 436} (1998) 257},
\href{http://arxiv.org/abs/hep-ph/9804398}{{\ttfamily arXiv:hep-ph/9804398
  [hep-ph]}}.

\bibitem{Dimopoulos:2001hw}
S.~Dimopoulos and G.~L. Landsberg,
  \href{http://dx.doi.org/10.1103/PhysRevLett.87.161602}{Phys. Rev. Lett.
  {\bfseries 87} (2001) 161602},
\href{http://arxiv.org/abs/hep-ph/0106295}{{\ttfamily arXiv:hep-ph/0106295
  [hep-ph]}}.

\bibitem{Giddings:2001bu}
S.~B. Giddings and S.~D. Thomas,
  \href{http://dx.doi.org/10.1103/PhysRevD.65.056010}{Phys. Rev. D {\bfseries
  65} (2002) 056010}, \href{http://arxiv.org/abs/hep-ph/0106219}{{\ttfamily
  arXiv:hep-ph/0106219 [hep-ph]}}.

\bibitem{Chatrchyan:2013xva}
{CMS Collaboration}, \href{http://dx.doi.org/10.1007/JHEP07(2013)178}{JHEP
  {\bfseries 1307} (2013) 178},
\href{http://arxiv.org/abs/1303.5338}{{\ttfamily arXiv:1303.5338 [hep-ex]}}.

\bibitem{ATLAS:2012pu}
{ATLAS Collaboration}, \href{http://dx.doi.org/10.1007/JHEP01(2013)029}{JHEP
  {\bfseries 1301} (2013) 029},
\href{http://arxiv.org/abs/1210.1718}{{\ttfamily arXiv:1210.1718 [hep-ex]}}.

\bibitem{Chatrchyan:2013qha}
{CMS Collaboration}, \href{http://dx.doi.org/10.1103/PhysRevD.87.114015}{Phys.
  Rev. D {\bfseries 87} (2013) 114015},
  \href{http://arxiv.org/abs/1302.4794}{{\ttfamily arXiv:1302.4794 [hep-ex]}}.

\bibitem{CMS:2012yf}
{CMS Collaboration}, \href{http://dx.doi.org/10.1007/JHEP01(2013)013}{JHEP
  {\bfseries 1301} (2013) 013},
  \href{http://arxiv.org/abs/1210.2387}{{\ttfamily arXiv:1210.2387 [hep-ex]}}.

\bibitem{Pythia:2007}
T.~Sjostrand, S.~Mrenna, and P.~Z. Skands,
  \href{http://dx.doi.org/10.1016/j.cpc.2008.01.036}{Comput. Phys. Commun.
  {\bfseries 178} (2008) 852}, \href{http://arxiv.org/abs/0710.3820}{{\ttfamily
  arXiv:0710.3820 [hep-ph]}}.

\bibitem{Gleisberg:2008ta}
T.~Gleisberg {et~al.},
  \href{http://dx.doi.org/10.1088/1126-6708/2009/02/007}{JHEP {\bfseries 0902}
  (2009) 007}, \href{http://arxiv.org/abs/0811.4622}{{\ttfamily arXiv:0811.4622
  [hep-ph]}}.

\bibitem{Pumplin:2002vw}
J.~Pumplin {et~al.},
  \href{http://dx.doi.org/10.1088/1126-6708/2002/07/012}{JHEP {\bfseries 0207}
  (2002) 012}, \href{http://arxiv.org/abs/hep-ph/0201195}{{\ttfamily
  arXiv:hep-ph/0201195 [hep-ph]}}.

\bibitem{Lai:2010vv}
H.-L. Lai {et~al.}, \href{http://dx.doi.org/10.1103/PhysRevD.82.074024}{Phys.
  Rev. D {\bfseries 82} (2010) 074024},
  \href{http://arxiv.org/abs/1007.2241}{{\ttfamily arXiv:1007.2241 [hep-ph]}}.

\bibitem{Gingrich:2009da}
D.~M. Gingrich, \href{http://dx.doi.org/10.1016/j.cpc.2010.07.027}{Comput.
  Phys. Commun. {\bfseries 181} (2010) 1917},
  \href{http://arxiv.org/abs/0911.5370}{{\ttfamily arXiv:0911.5370 [hep-ph]}}.

\bibitem{Martin:2009iq}
A.~Martin, W.~Stirling, R.~Thorne, and G.~Watt,
  \href{http://dx.doi.org/10.1140/epjc/s10052-009-1072-5}{Eur. Phys. J. C
  {\bfseries 63} (2009) 189}, \href{http://arxiv.org/abs/0901.0002}{{\ttfamily
  arXiv:0901.0002 [hep-ph]}}.

\bibitem{atlas:pythia8tunes}
{ATLAS Collaboration}, Tech. Rep. ATL-PHYS-PUB-2012-003, CERN, Geneva, 2012.
\newblock \url{http://cds.cern.ch/record/1474107/}.

\bibitem{Agostinelli2003250}
{GEANT4} Collaboration, S.~Agostinelli {et~al.},
  \href{http://dx.doi.org/10.1016/S0168-9002(03)01368-8}{Nucl. Instrum. Meth. A
  {\bfseries 506} (2003) 250}.

\bibitem{ATLASSimInfra}
{ATLAS Collaboration},
  \href{http://dx.doi.org/10.1140/epjc/s10052-010-1429-9}{Eur. Phys. J. C
  {\bfseries 70} (2010) 823}, \href{http://arxiv.org/abs/1005.4568}{{\ttfamily
  arXiv:1005.4568 [physics.ins-det]}}.

\bibitem{1748-0221-3-08-S08003}
{ATLAS Collaboration},
  \href{http://dx.doi.org/10.1088/1748-0221/3/08/S08003}{JINST {\bfseries 3}
  (2008) S08003}.

\bibitem{Aad:2012xs}
{ATLAS Collaboration},
  \href{http://dx.doi.org/10.1140/epjc/s10052-011-1849-1}{Eur. Phys. J. C
  {\bfseries 72} (2012) 1849}, \href{http://arxiv.org/abs/1110.1530}{{\ttfamily
  arXiv:1110.1530 [hep-ex]}}.

\bibitem{ref:lumi}
{ATLAS Collaboration},
  \href{http://dx.doi.org/10.1140/epjc/s10052-013-2518-3}{Eur. Phys. J. C
  {\bfseries 73} no.~8, (2013) 1},
  \href{http://arxiv.org/abs/1302.4393}{{\ttfamily arXiv:1302.4393 [hep-ex]}}.

\bibitem{springerlink:10.1140/epjc/s10052-010-1512-2}
{ATLAS Collaboration},
  \href{http://dx.doi.org/10.1140/epjc/s10052-010-1512-2}{Eur. Phys. J. C
  {\bfseries 71} (2011) 1512}, \href{http://arxiv.org/abs/1009.5908}{{\ttfamily
  arXiv:1009.5908 [hep-ex]}}.

\bibitem{Cacciari:2008gp}
M.~Cacciari, G.~P. Salam, and G.~Soyez,
  \href{http://dx.doi.org/10.1088/1126-6708/2008/04/063}{JHEP {\bfseries 0804}
  (2008) 063},
\href{http://arxiv.org/abs/0802.1189}{{\ttfamily arXiv:0802.1189 [hep-ph]}}.

\bibitem{Cacciari:2008gn}
M.~Cacciari, G.~P. Salam, and G.~Soyez,
  \href{http://dx.doi.org/10.1088/1126-6708/2008/04/005}{JHEP {\bfseries 0804}
  (2008) 005},
\href{http://arxiv.org/abs/0802.1188}{{\ttfamily arXiv:0802.1188 [hep-ph]}}.

\bibitem{Cacciari:2009dp}
M.~Cacciari, G.~P. Salam, and S.~Sapeta,
  \href{http://dx.doi.org/10.1007/JHEP04(2010)065}{JHEP {\bfseries 1004} (2010)
  065},
\href{http://arxiv.org/abs/0912.4926}{{\ttfamily arXiv:0912.4926 [hep-ph]}}.

\bibitem{2010JES}
{ATLAS Collaboration},
  \href{http://dx.doi.org/10.1140/epjc/s10052-013-2304-2}{Eur. Phys. J. C
  {\bfseries 73} no.~3, (2013) 2304},
  \href{http://arxiv.org/abs/1112.6426}{{\ttfamily arXiv:1112.6426 [hep-ex]}}.

\bibitem{Aad:2011tw}
{ATLAS Collaboration},
  \href{http://dx.doi.org/10.1016/j.physletb.2011.11.010}{Phys. Lett. B
  {\bfseries 706} (2011) 150}, \href{http://arxiv.org/abs/1108.0253}{{\ttfamily
  arXiv:1108.0253 [hep-ex]}}.

\bibitem{PhysRevD.83.052005}
{ATLAS Collaboration},
  \href{http://dx.doi.org/10.1103/PhysRevD.83.052005}{Phys. Rev. D {\bfseries
  83} (2011) 052005}, \href{http://arxiv.org/abs/1012.4389}{{\ttfamily
  arXiv:1012.4389 [hep-ex]}}.

\bibitem{Harris:2011bh}
R.~M. Harris and K.~Kousouris,
  \href{http://dx.doi.org/10.1142/S0217751X11054905}{Int. J. Mod. Phys. A
  {\bfseries 26} (2011) 5005},
\href{http://arxiv.org/abs/1110.5302}{{\ttfamily arXiv:1110.5302 [hep-ex]}}.

\bibitem{Choudalakis11b}
G.~Choudalakis and D.~Casadei,
  \href{http://dx.doi.org/10.1140/epjp/i2012-12025-y}{Eur. Phys. J. Plus
  {\bfseries 127} (2012) 15}, \href{http://arxiv.org/abs/1111.2062}{{\ttfamily
  arXiv:1111.2062 [physics.data-an]}}.

\bibitem{Choudalakis11a}
G.~Choudalakis, \href{http://arxiv.org/abs/1101.0390}{{\ttfamily
  arXiv:1101.0390 [physics.data-an]}}.

\end{thebibliography}\endgroup

\clearpage
\begin{flushleft}
{\Large The ATLAS Collaboration}

\bigskip

G.~Aad$^{\rm 48}$,
T.~Abajyan$^{\rm 21}$,
B.~Abbott$^{\rm 112}$,
J.~Abdallah$^{\rm 12}$,
S.~Abdel~Khalek$^{\rm 116}$,
O.~Abdinov$^{\rm 11}$,
R.~Aben$^{\rm 106}$,
B.~Abi$^{\rm 113}$,
M.~Abolins$^{\rm 89}$,
O.S.~AbouZeid$^{\rm 159}$
% ...
\bigskip
\\
% ...
$^{11}$ Institute of Physics, Azerbaijan Academy of Sciences, Baku, Azerbaijan\\
$^{12}$ Institut de F{\'\i}sica d'Altes Energies and Departament de F{\'\i}sica de la Universitat Aut{\`o}noma de Barcelona, Barcelona, Spain\\
% ...
$^{21}$ Physikalisches Institut, University of Bonn, Bonn, Germany\\
% ...
$^{48}$ Fakult{\"a}t f{\"u}r Mathematik und Physik, Albert-Ludwigs-Universit{\"a}t, Freiburg, Germany\\
% ...
$^{89}$ Department of Physics and Astronomy, Michigan State University, East Lansing MI, United States of America\\
% ...
$^{106}$ Nikhef National Institute for Subatomic Physics and University of Amsterdam, Amsterdam, Netherlands\\
% ...
$^{112}$ Homer L. Dodge Department of Physics and Astronomy, University of Oklahoma, Norman OK, United States of America\\
$^{113}$ Department of Physics, Oklahoma State University, Stillwater OK, United States of America\\
% ...
$^{116}$ LAL, Universit{\'e} Paris-Sud and CNRS/IN2P3, Orsay, France\\
% ...
$^{159}$ Department of Physics, University of Toronto, Toronto ON, Canada\\
% ...
\end{flushleft}


\end{document}